\documentstyle{article}
\input epsf
\newcommand{\beq}{\begin{equation}}
\newcommand{\eeq}{\end{equation}}

\title{Kink dynamics in a novel discrete sine-Gordon system}
\author{J M Speight\thanks{Current e-mail address:
{\tt speight@amsta.leeds.ac.uk}}
\ and \ R S Ward \\
        Department of Mathematical Sciences, \\
        University of Durham, Durham DH1 3LE}
\date{}

\begin{document}

\maketitle

\begin{abstract}

A spatially-discrete sine-Gordon system with some novel features is described.
There is a topological or Bogomol'nyi lower bound on the energy of a kink,
and an explicit static kink which saturates this bound. There is no Peierls
potential barrier, and consequently the motion of a kink is simpler,
especially at low speeds. At higher speeds, it radiates and slows down.
\\
\\ AMS Classification numbers 58F, 70F.
\end{abstract}

\section{Introduction}

There are many nonlinear systems, in one spatial dimension, which
admit topologically-stable kink solutions: for example, the sine-Gordon
and phi-four systems. They have many physical applications. For applications
in, say, condensed-matter physics or biophysics, an accurate model should
take the discreteness of space into account --- in other words, the kinks
live on a one-dimensional lattice rather than in a continuum. A discrete
version of the sine-Gordon system which has been extensively studied 
is the Frenkel-Kontorova model, in which the partial derivative
$\partial/\partial x$ of the continuum system 
is replaced by the forward difference
on the lattice. In this case, much of the topological character of the model
is lost, and there is no known explicit kink solution. 

Let us briefly recall two of the features of kink motion in the 
Frenkel-Kontorova model [1--5].
 First, the energy of a kink depends on its position
in relation to the lattice: a static kink located halfway between two lattice
sites has a lower energy than one located exactly on a lattice site. The
difference between these two energies is the so-called Peierls-Nabarro
barrier. So if a moving kink does not have enough kinetic energy, then it
gets trapped between two adjacent lattice sites; it oscillates, and emits
radiation, and cannot escape. Secondly, a kink which starts off with enough
speed to surmount the Peierls-Nabarro barrier, and move along the lattice,
will lose energy (as radiation) and slow down, until eventually it too
becomes trapped. 

In this note, we wish to describe an alternative discrete sine-Gordon
system, in which the Peierls-Nabarro barrier is eliminated. It maintains
an important feature of the continuum model, namely a topological lower
bound (``Bogomol'nyi bound'') on the energy of a kink. And it
admits an explicit static kink solution on the lattice, which saturates
this lower bound. A moving kink will still radiate and slow down, but it is
never trapped, no matter how slowly it moves. So this lattice sine-Gordon 
model is quite different from the usual one described above.

In the next section we shall describe the model. Subsequent sections study
kink motion, for low speed and for (moderately) high speed. In each case,
we make approximate analytic predictions of the kink behaviour, and
compare these with  numerical simulations.

\section{A topological discrete sine-Gordon system} 

Let us begin with a very brief review of the situation for the continuous
sine-Gordon equation $\varphi_{xx}-\varphi_{tt} = \sin\varphi$.
It is convenient to use the dependent variable
$\psi=\frac{1}{2}\varphi$ instead of $\varphi$. The potential
energy of the field is
\begin{equation}
E_{P}=\frac{1}{4}\int_{-\infty}^{\infty} (\psi_{x}^{2} + \sin^{2}\psi)\, dx  
\end{equation}
(this is normalized so that a single static kink has unit energy).
The kink boundary condition is $\psi\to 0$ as $x\to-\infty$, $\psi\to\pi$
as $x\to\infty$. The standard Bogomol'nyi argument \cite{B} is
\begin{eqnarray}
0 &\leq& \frac{1}{4}\int_{-\infty}^{\infty}(\psi_{x}-\sin\psi)^{2}\,dx   
               \nonumber \\
  &=& E_{P}+\frac{1}{2}\int_{-\infty}^{\infty}\partial_{x}(\cos\psi)\,dx   
               \nonumber \\
  &=& E_{P}-1, 
\end{eqnarray}
where kink boundary conditions are imposed. So the energy $E_{P}$ is bounded
below by 1; and $E_{P}$ equals 1 if and only if $\psi_{x}=\sin\psi$, the
solution of which is the static kink
\begin{equation}
\psi(x) = 2\arctan\exp(x-x_{0}).
\end{equation}

From this point on, $x$ becomes a discrete variable, with lattice spacing
$h$. The subscript + denotes forward shift, ie. $f_{+}(x)=f(x+h)$; so, for
example, the forward difference is given by $\Delta f=h^{-1}(f_{+} -f)$.
To obtain a lattice version of the Bogomol'nyi bound, we may begin with
the same function $\cos\psi$ as appears in (2), and
``reconstruct'' the inequality. The first step is to choose a
factorization
\begin{equation}
\Delta\cos\psi = -DF,
\end{equation}
where $D\to\psi_{x}$ and $F\to\sin\psi$ in the continuum limit $h\to 0$.
Then define the potential energy of the lattice sine-Gordon field to be
\begin{equation}
E_{P} = \frac{h}{4} \sum_{x=-\infty}^{\infty}(D^{2}+F^{2}). 
\end{equation}
It follows, just as in the continuum case, that (with kink boundary
conditions) $E_{P}$ is bounded below by 1; and the minimum is attained
if and only if $D=F$.

There is a choice involved in (4), and we are free to make whatever choice
we like. The most natural one seems to be
\begin{eqnarray}
D &=& \frac{2}{h}\sin\frac{1}{2}(\psi_{+}-\psi), \nonumber \\
F &=& \sin\frac{1}{2}(\psi_{+}+\psi).
\end{eqnarray}
Substituting these into (5) gives
\begin{equation}
E_{P}=\frac{h}{4}\sum_{x=-\infty}^{\infty}[\frac{4}{h^{2}}\sin^{2}
      \frac{1}{2}(\psi_{+}-\psi) + \sin^{2}\frac{1}{2}(\psi_{+} +\psi)],
\end{equation}           
which, of course, reduces to (1) in the continuum limit.

Let us summarize what we have so far. The real-valued field $\psi(t,x)$
depends on the continuous variable $t$ and the discrete variable $x$.
Its potential energy $E_{P}$ is defined by the expression (7); for the kinetic
energy, we may for example make the simple choice
\begin{equation}
E_{K}=\frac{h}{4} \sum_{x=-\infty}^{\infty} \dot{\psi}^{2}, 
\end{equation}
where $\dot{\psi}=d\psi/ dt$. 
The boundary condition on $\psi$ is that it
should tend to an integer multiple of $\pi$ (independent of time) as
$x\to-\infty$ or $x\to+\infty$; this guarantees finite energy. A 1-kink
configuration has the boundary values $\psi\to 0$ as $x\to-\infty$ and
$\psi\to\pi$ as $x\to+\infty$. For such fields, the total energy
$E=E_{P}+E_{K}$ is bounded below by 1; and this lower bound is attained if
and only if $\dot{\psi}=0$, and the two expressions in (6) are equal.

 This 
latter condition, namely $D=F$, is called the Bogomol'nyi equation \cite{B}.
It is a first-order difference equation, whose solutions (for kink boundary
conditions) minimize the potential energy. Hence these (static) solutions
are also solutions of the Euler-Lagrange equations
\begin{equation}
\ddot{\psi} = -\frac{2}{h}\frac{\partial E_{P}}{\partial \psi},
\end{equation}
since $\partial E_{P}/\partial \psi=0$ at a minimum. This is a general feature
of the Bogomol'nyi argument: one gets first-order equations whose solutions
are also static solutions of the second-order equations of motion.
Furthermore (and this is the most  important feature), solutions of the
Bogomol'nyi equations have an energy which is at its topological minimum
value.

In the expression (7) for $E_{P}$, the first term in the summand represents
an attractive force (nonlinear Hooke's law) which couples nearest
neighbours, and the second term a substrate potential depending on the
average of pairs of nearest neighbours. The strengths of these two forces
have, in effect, been normalized by scaling, and $h$ may be regarded as a
dimensionless parameter in the model.

The Bogomol'nyi equation $D=F$ may also be written as
\begin{equation} 
\tan\frac{\psi_{+}}{2} = \frac{2+h}{2-h}\tan\frac{\psi}{2}, 
\end{equation}
from which one sees that the parameter $h$ should be less than 2 if
one is to obtain a well-behaved solution. In fact, the solution of (10) can
be written down explicitly: given that $0<h<2$, it is
\begin{equation} 
\psi = 2\arctan\exp a(x-b),
\end{equation}
where
\begin{equation}
a=\frac{2}{h}{\rm arctanh}\frac{h}{2}, 
\end{equation}
and where $b$ is an arbitrary real constant. This is a static
lattice kink solution. Because it corresponds to a minimum of the energy in the
kink sector, it is stable under perturbations which remain in that sector; in
other words, under perturbations which preserve the kink boundary conditions on
$\psi$. The location of the kink is determined by the continuous
parameter (zero-mode) $b$. The kink is highly localized, in the sense that its
energy density (the summand in eqn 7) is concentrated on just a couple of
links of the lattice. For example, if $b=0$, then the proportion of the
energy contributed by the two links $[-h,0]$ and $[0,h]$ is $4h/(4+h^{2})$;
if $h=1$ this proportion is 80\%, and as $h\to2$ it tends to 100\%.
A diagram illustrating the kink solution $\psi(x)$ for $h=1$ is presented in
figure 1.

It is instructive to look at the limit $h=2$. In this case, the solution of
the Bogomol'nyi equation is supported at a single lattice site $x_{0}$:
\begin{equation}
\psi(x) = \left\{ \begin{array}{lll}
                      0    & \mbox{if $x<x_{0}$} \\
                      c    & \mbox{if $x=x_{0}$} \\
                      \pi  & \mbox{if $x>x_{0}$}
                      \end{array}
             \right. 
\end{equation}                          
where $c$ is a real constant. By contrast with (11), the
``position''\ zero-mode is discrete; and there is an additional
continuous zero-mode, namely the value $c$ of $\psi$ at the centre of the
kink. This limiting case is therefore marginally unstable: a perturbation will in
general cause $\psi(x_{0})$ to increase or decrease without bound.

\vbox{
\centerline{\epsfysize=3truein
\epsfbox[63 420 544 745]{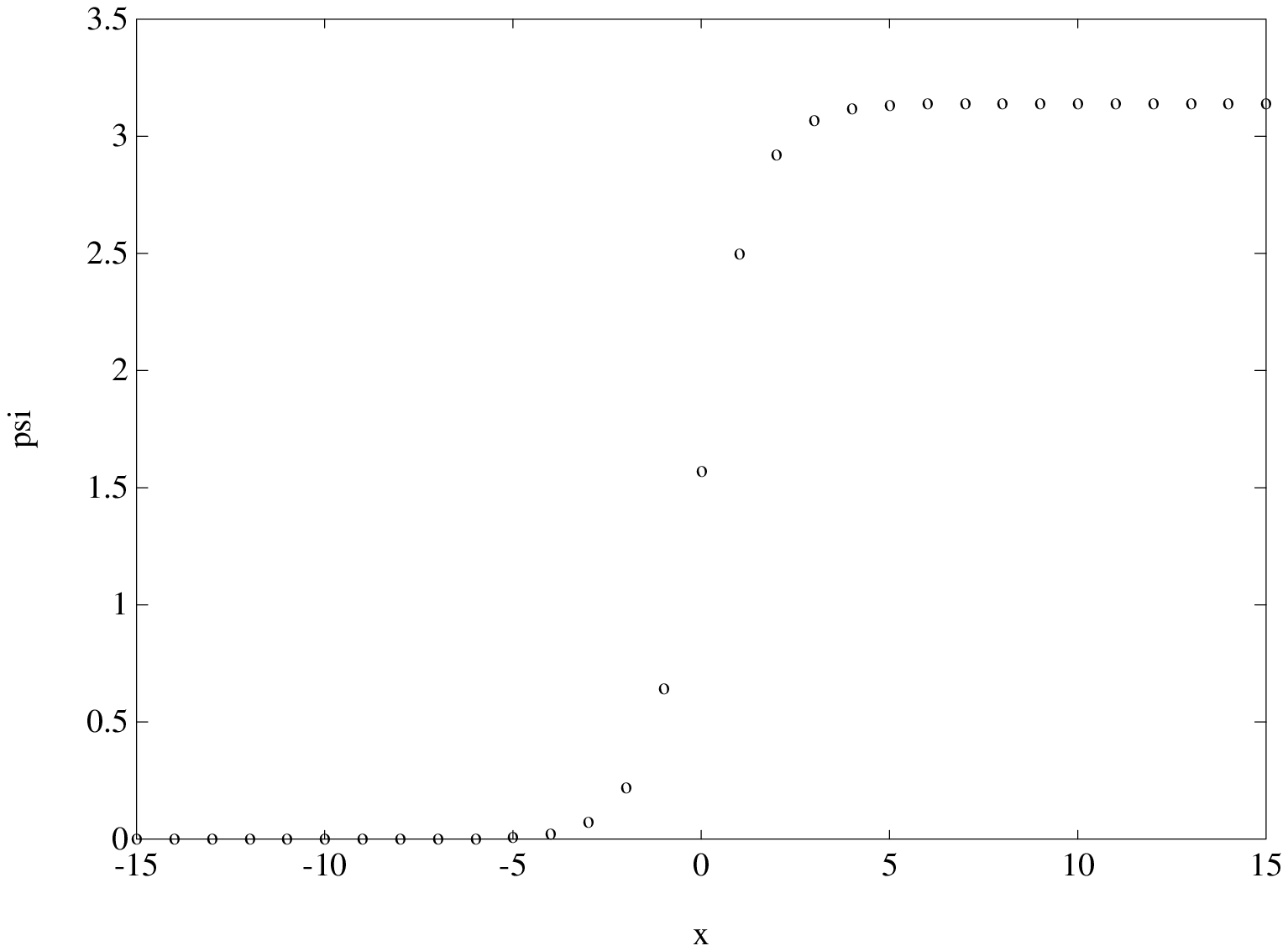}}
\centerline{\it Figure 1: The static kink for $h=1$.}
}
\vspace{0.5cm}

The situation we wish to study is that of the moving kink. There is no explicit
solution in this case, and so one has to resort to approximation, or to
numerical solution of the equations of motion (9), namely
\begin{eqnarray}
\ddot{\psi} &=& \frac{1}{h^{2}}[\sin(\psi_{+}-\psi)-\sin(\psi-\psi_{-})]
     -\frac{1}{4}[\sin(\psi_{+}+\psi)+\sin(\psi+\psi_{-})] \nonumber \\
            &=& \frac{4-h^{2}}{4h^{2}}\cos\psi(\sin\psi_{+}+\sin\psi_{-})
               -\frac{4+h^{2}}{4h^{2}}\sin\psi(\cos\psi_{+}+\cos\psi_{-}).
\end{eqnarray}

\section{A collective-coordinate approximation}
 
The simplest approximation of kink dynamics is 
obtained by restricting $\psi$ to have the form (11),
with $b$ now becoming a dynamical variable $b(t)$. So the number of degrees of 
freedom is reduced from infinity to one. Our conjecture is that this 
appproximation is a good one, provided that the speed $v$ of the kink is small
(cf \cite{M1,M2}); we shall elaborate on this condition later in the section.
So the Lagrangian is
\begin{eqnarray}
L &=& E_{K}-E_{P} \nonumber \\
  &=& f(b)\dot{b}^{2}-1,
\end{eqnarray}
where
\begin{equation}
\label{16}
f(b)=\frac{a^{2}h}{4}\sum_{x=-\infty}^{\infty}{\rm sech}^{2}a(x-b).
\end{equation}
Note that $f(b)$ is periodic with period $h$, is even, and has a local maximum at
$b=0$. Graphs of $f(b)$ for various values of $h$ are given in figure 2.

The Euler-Lagrange equation of the system is
\begin{equation}
f(b) \ddot{b} =-f'(b)\dot{b}^{2}.
\end{equation}
This may be reduced to quadratures:
\begin{eqnarray}
   vt & = & \int_{0}^{b(t)}\sqrt{\frac{f(\tilde{b})}{f(0)}}\,d\tilde{b}   
             \nonumber \\
    & \equiv & F_{h}(b),
\end{eqnarray}
where $b(0)=0$, $\dot{b}(0)=v$. The function $F_{h}(b)$ is strictly increasing,
and satisfies
\begin{equation}                                                             
F_{h}(b+nh)=F_{h}(b)+nF_{h}(h)
\end{equation}
for all integers $n$ owing to the
periodicity of $f(b)$, so it suffices to know $F_{h}(b)$ for
$0 \leq b<h$.

\vbox{
\centerline{\epsfysize=3truein
\epsfbox[63 420 544 745]{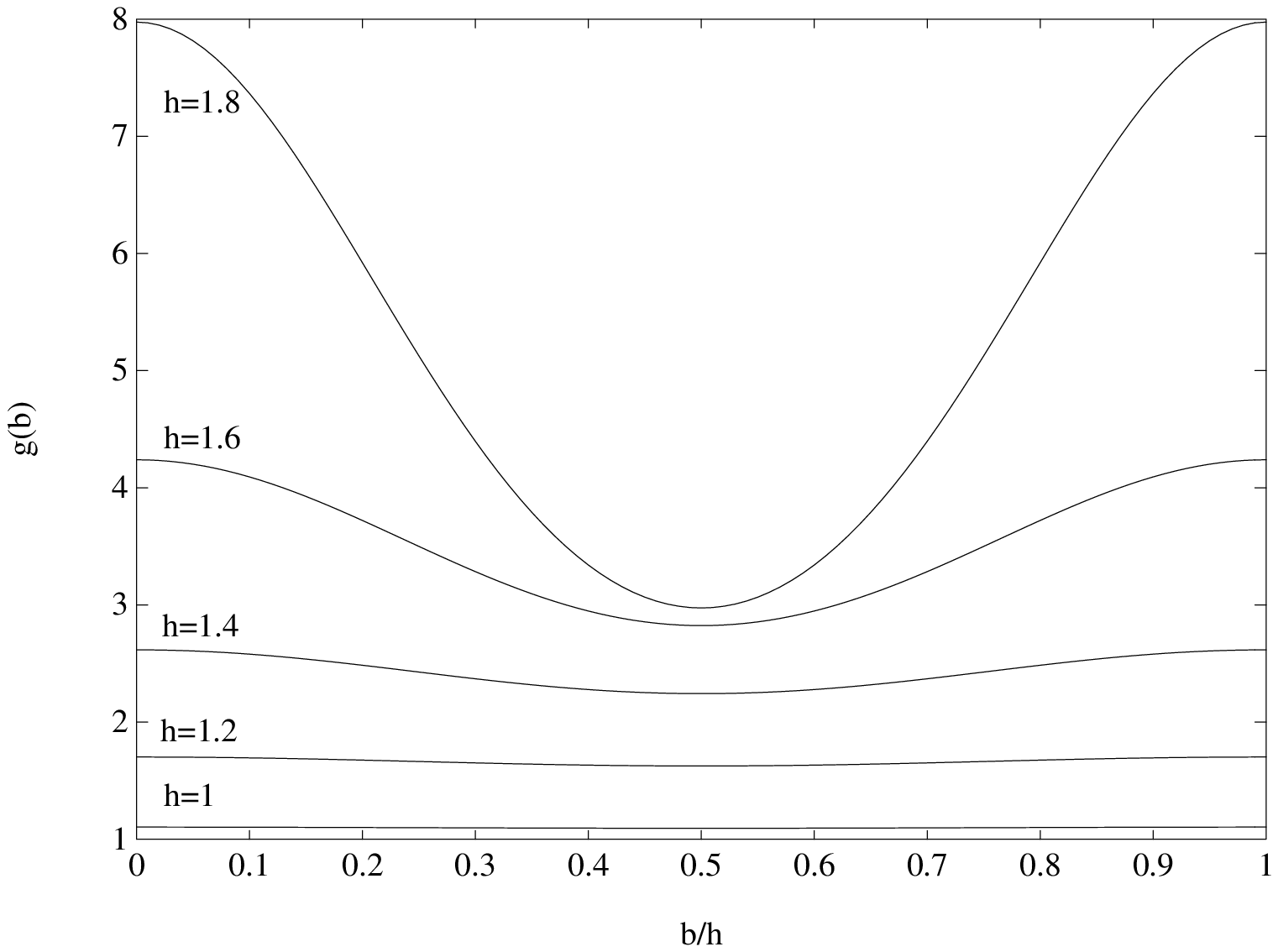}}
\centerline{\it Figure 2: The function $f(b)$ of equation (\ref{16}).}
}
\vspace{0.5cm}

\vbox{
\centerline{\epsfysize=3truein
\epsfbox[62 420 544 745]{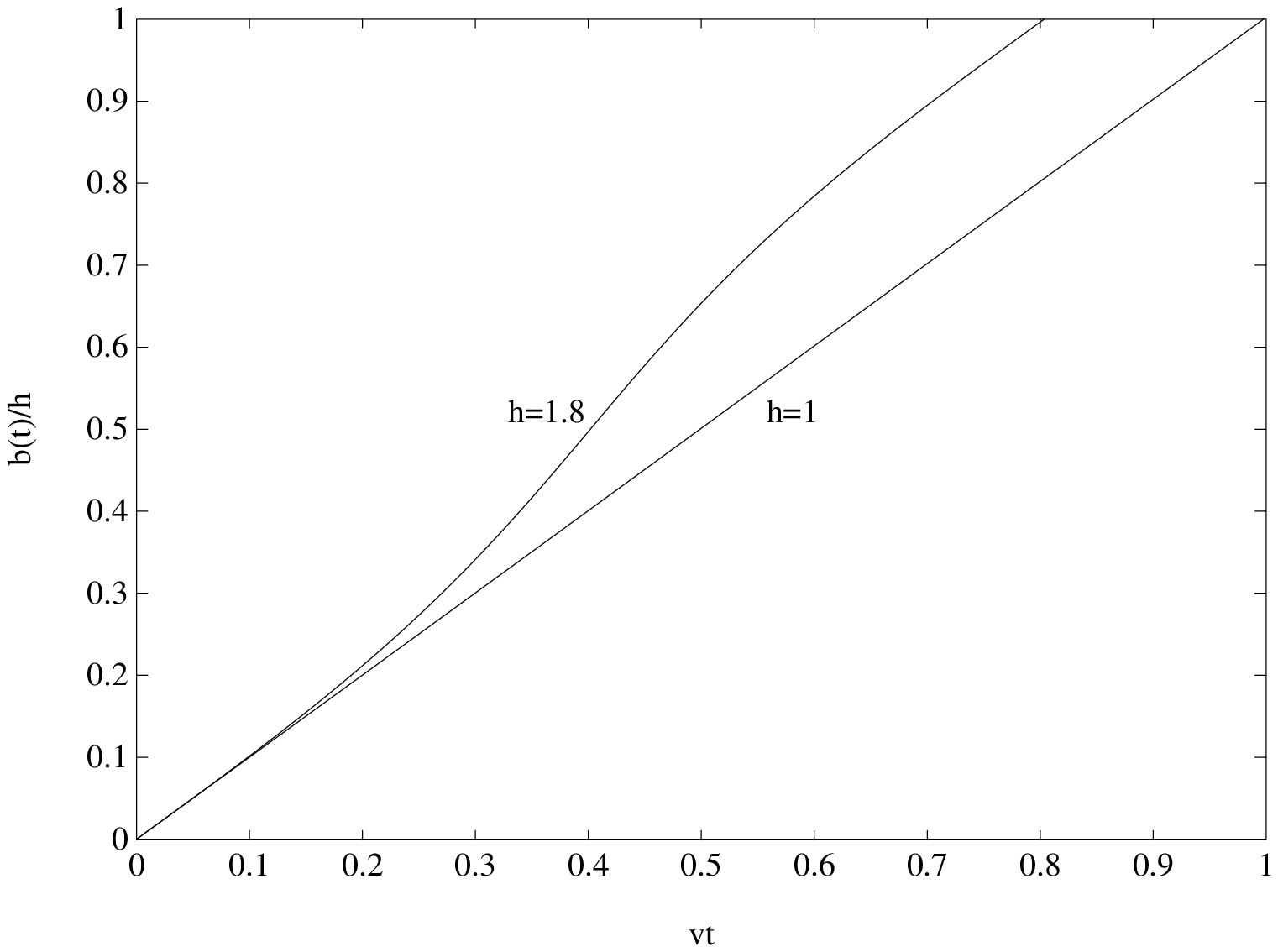}}
\centerline{\it Figure 3: The function $b(t)$ over one wobble period for
$h=1$ and $h=1.8$.}
}
\vspace{0.5cm}

This function $F_{h}$ is easily inverted (fig.\ 3) to give the kink trajectory
\begin{equation}
b(t)=F_{h}^{-1}(vt).
\end{equation}
 The main feature is that the kink wobbles 
as it moves through the lattice. This wobble is a dynamical effect; there is of
course no static potential (such as Peierls-Nabarro) causing it.
The time taken for the kink to move from $x=0$ to $x=h$ (the wobble period) is
\begin{equation}
T=\frac{F_{h}(h)}{v};
\end{equation}
 and $F_{h}(h)/h<1$, approaching the upper bound
in the continuum limit, $h\rightarrow0$.                         
That is, the kink travels faster by a factor $h/F_{h}(h)$ than would be
na\"{\i}vely suggested by the initial velocity $v$.
A graph of $F_{h}(h)$
(fig.\ 4) shows that discreteness effects are small for $h<1.2$,
but grow large as $h$ approaches 2.

We believe that this approximation of kink motion is accurate for small $v$.
The Bogomol'nyi bound is crucial to this belief: the idea is that the
configuration sits at the bottom of a one-dimensional potential valley, and
the kink can move freely along the valley floor (ie. in the direction of the
zero-mode $b$). There are at least two ways in which the approximation differs
from the full model. First, there is dynamical dressing (a moving kink has a
different shape from a static one). In the continuum limit, this correction is
of order $v^{2}$ (relativistic contraction), and one might expect the dressing
in the discrete case also to be of order $v^{2}$ (bearing in mind that the 
expression for the static kink is exact). Secondly, there is transfer
of energy to radiation (phonons). In Bogomol'nyi field theories, the
suggestion is that this energy transfer will be suppressed by a factor of order
$\exp(-T_{0}/T_{1})$, where $T_{0}$ is a typical timescale for the truncated
dynamics, and $T_{1}$ is the maximum period of the radiation in the theory
\cite{M2}. Our suggestion (not proved) is that the same is true in Bogomol'nyi 
lattice models such as the present one. In this case, $T_{0}$ is the wobble 
period (21), which is of order $h/v$; and $T_{1}=2\pi$, since the minimum phonon
frequency is unity (see next section). So the energy transfer (per lattice
site traverse) is suppressed by $\exp[-h/(2\pi v)]$, and is therefore
negligible for small $v$ (as long as $h$ is not small). If $h$ {\em is} small,
in fact if $h$ is less than $1$, then the amplitude of the wobble is tiny,
and so one would expect the radiation to be negligible irrespective of $v$;
of course, this is exactly what happens in the continuum ($h\rightarrow0$)
limit.

\vbox{
\centerline{\epsfysize=3truein
\epsfbox[63 420 544 745]{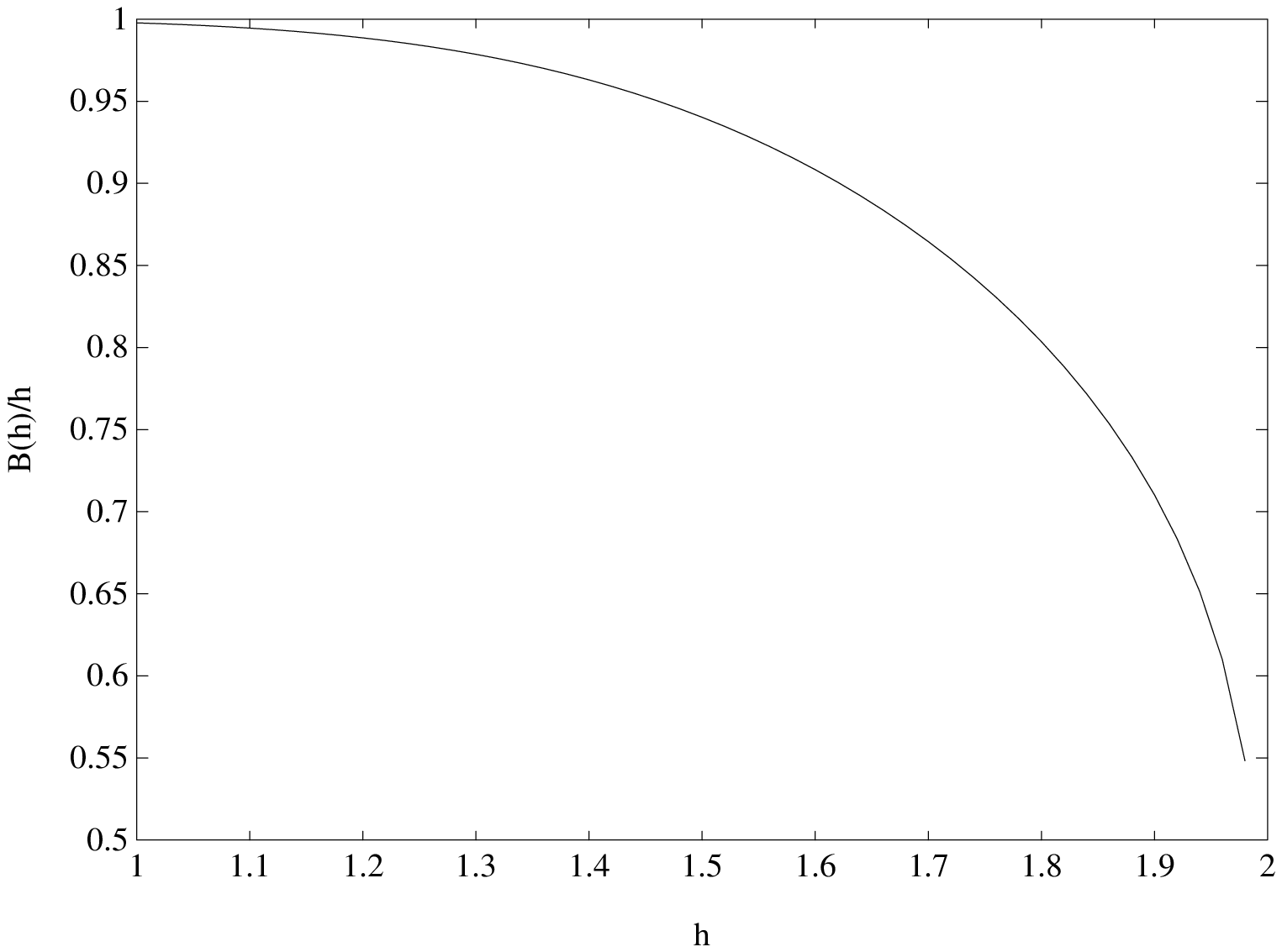}}
\centerline{\it Figure 4: Effect of discreteness on the function 
$F_{h}(h)/h$.}
}
\vspace{0.5cm}

The accuracy of the approximation has been tested numerically using a
fully-explicit fourth-order Runge-Kutta algorithm with fixed time-step
$0.01$. The initial condition was a Galilean-boosted static kink
profile with initial velocity $0.01$ lattice sites per unit time  
($v=0.01h$). Simulations of duration 1000 time units were performed for
$h=1, 1.2, 1.4, 1.6$ and $1.8$. In every case the kink moves freely,
without pinning, undergoing motion of the predicted periodicity.
Furthermore, inspection of the kink velocity over a single period
reveals close agreement with $\dot{b}(t)$ calculated from (20)
(see fig.\ 5), although for $h=1.2$, numerical errors rather swamp the
very small theoretical wobble (note the scale on the velocity axis).

Collective-variable analyses of the Frenkel-Kontorova model (cf \cite{BWE}) have
introduced the kink position as an extra variable, accompanied by a constraint.
But since the dynamics is much more complicated, all the degrees of freedom have
to be kept in order to obtain accurate results. A truncation to one degree of
freedom, as in the present case, does not work.

\vbox{
\centerline{\epsfysize=3truein
\epsfbox[63 420 544 745]{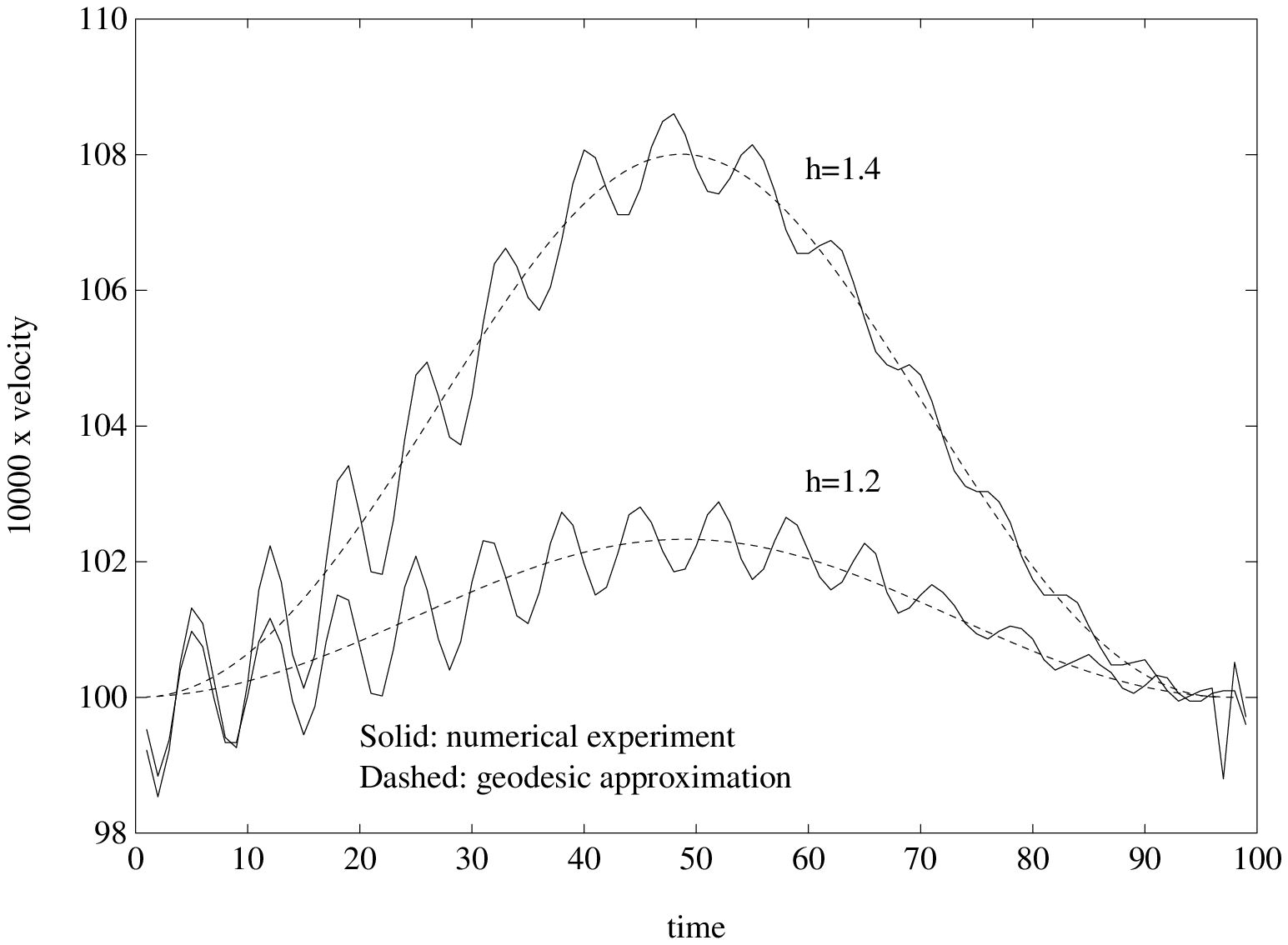}}
}

\vbox{
\centerline{\epsfysize=3truein
\epsfbox[63 420 544 745]{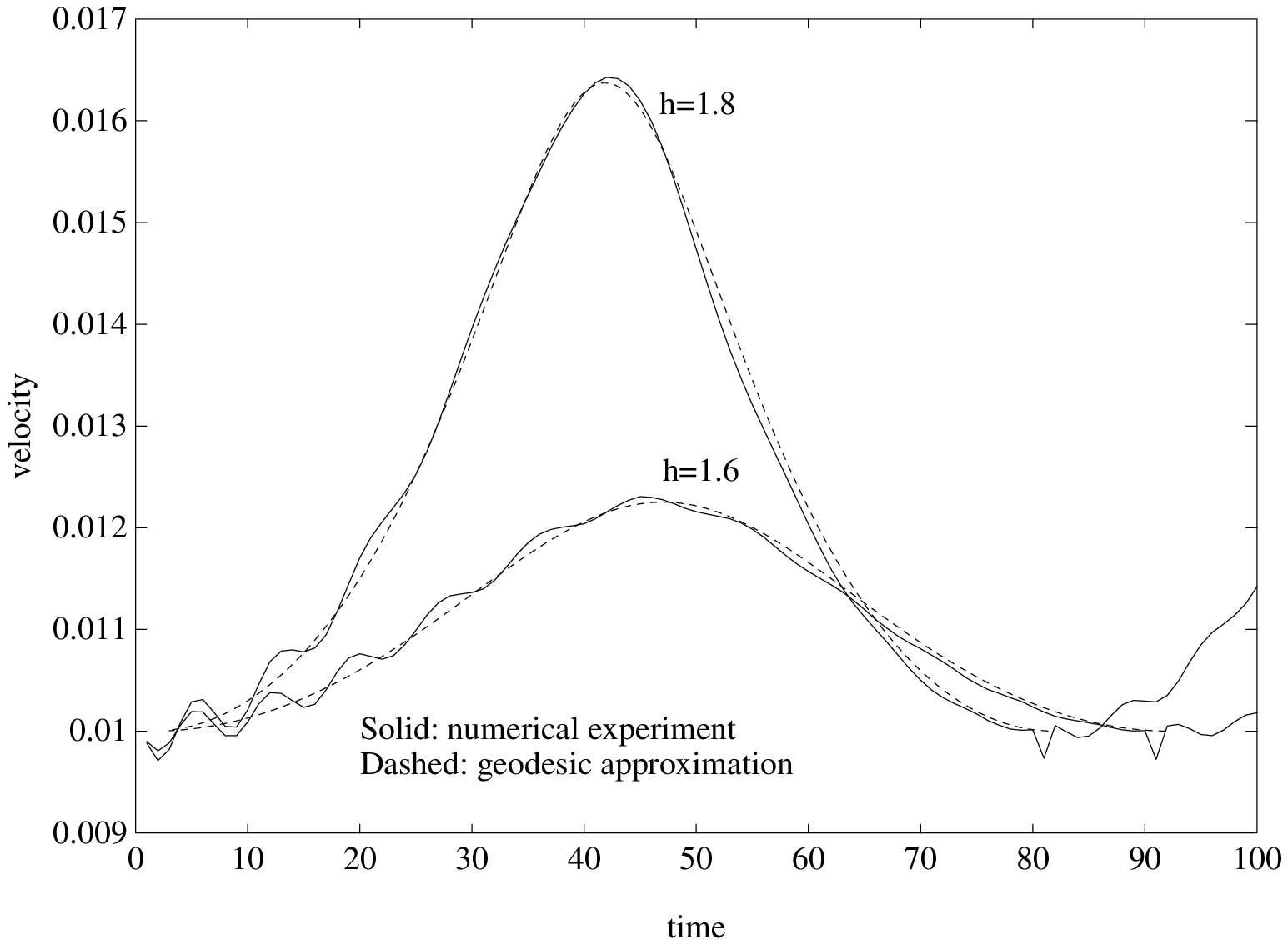}}
\centerline{\it Figure 5: Kink velocity over a single wobble period.}
}

\section{Fast-moving kinks}

The collective-coordinate approximation is expected to fail at high           
velocities (except for small $h$).
 This is observed in the simulations as a gradual onset of
kink deceleration as the initial velocity is increased. Failure occurs
at lower velocities for coarser lattices---around $v=0.012$ for $h=1.8$
compared with $v=0.15$ for $h=1$. The kink energy is dissipated in the form of
small amplitude oscillations (``radiation'' or ``phonons'') emitted in its
wake, propagating backwards (see fig.\ 6).

\vbox{
\centerline{\epsfysize=3truein
\epsfbox[550 50 1330 570]{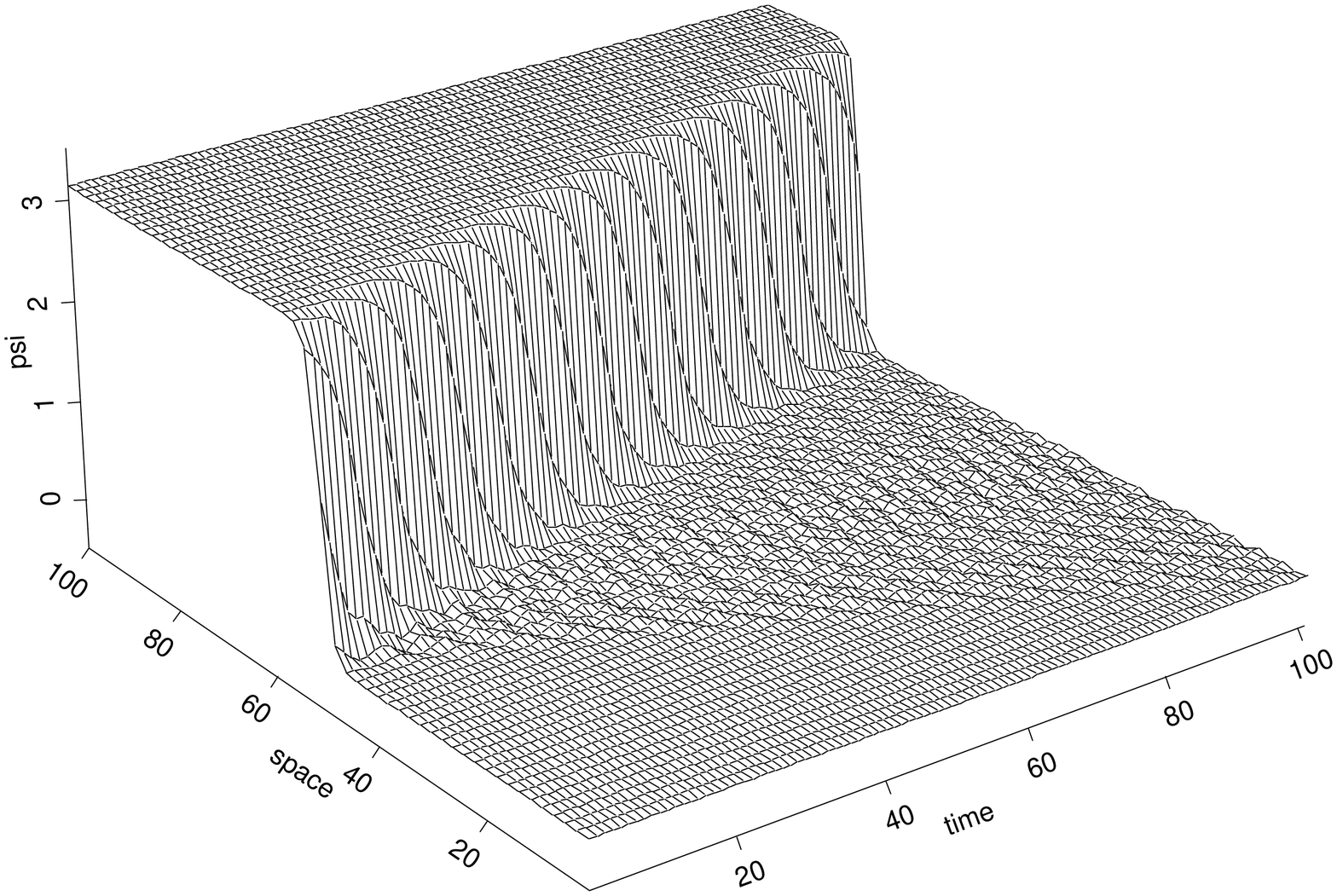}}
\centerline{\it Figure 6: Radiation by a fast-moving kink.}
}
\vspace{0.5cm}

The effect of this radiation on the kink velocity over a long time-scale
can be seen in fig.\ 7. The data were produced by the above-mentioned 
Runge-Kutta algorithm, run for $16000$ time units with a time-step of $0.01$.
The initial configuration was a static kink Galilean-boosted to speed
$v=0.3$, on a lattice of unit spacing ($h=1$). To cut reflexion of radiation 
from the fixed left-hand boundary, the first five lattice sites were damped.
After an initial velocity drop of $0.02$ in $10$ time units as the kink
assumes a more appropriate shape, it decelerates more slowly. The modulation 
of the amplitude of velocity oscillations is due to the velocity sampling
(once every 10 time units) falling in and out of phase with the periodic wobble
of the kink as it passes lattice sites.

\vspace{0.5cm}
\vbox{
\centerline{\epsfysize=3truein
\epsfbox[63 420 544 745]{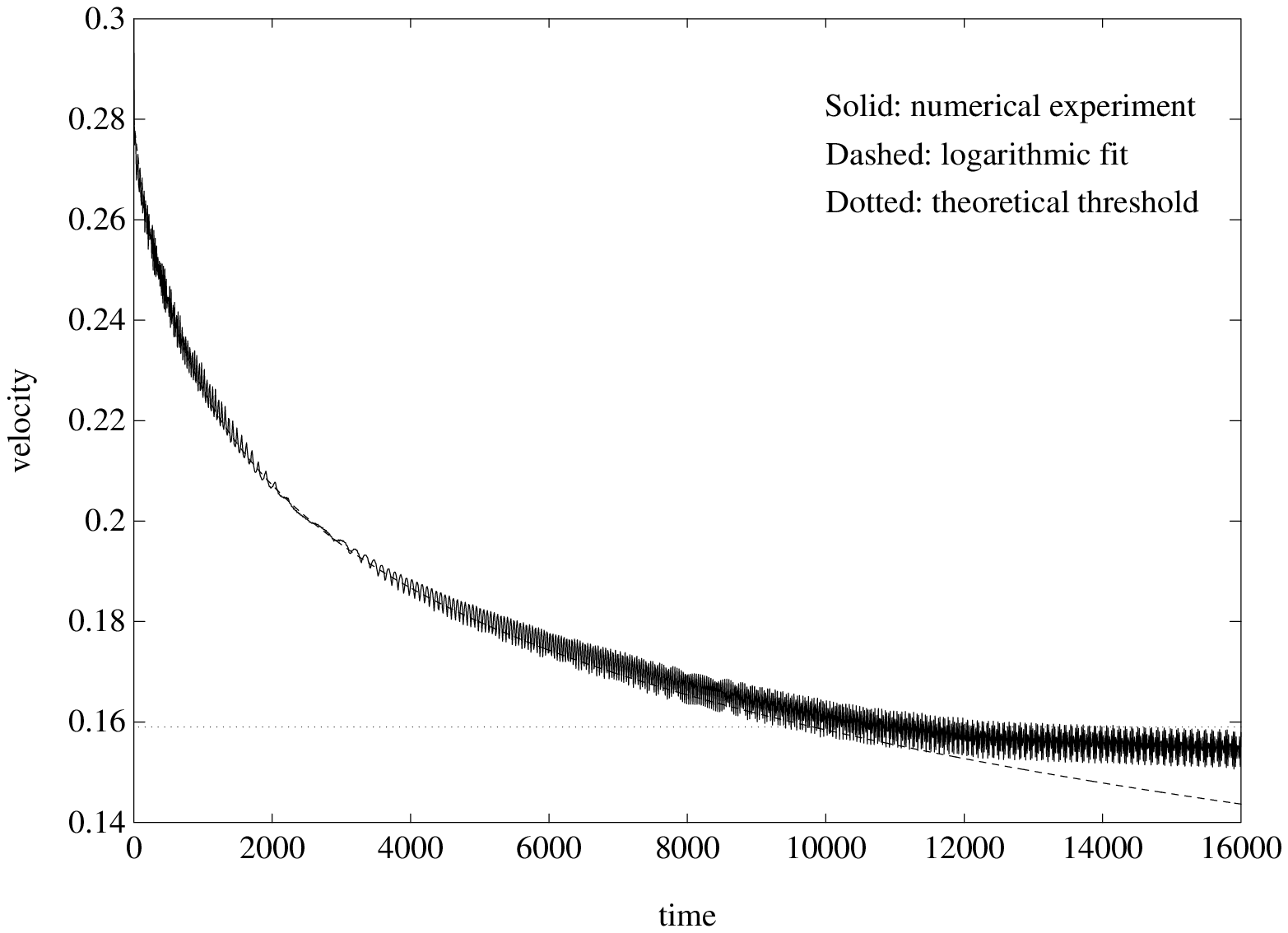}}
\centerline{\it Figure 7: Radiative kink deceleration.}
}

\newpage
The most interesting feature of fig.\ 7 is the existence of a threshold 
velocity, $v \approx 0.16$, below which deceleration, and hence
radiation, is much reduced. Some understanding of this phenomenon may be
gained by an analysis (motivated by \cite{PK}) of the linearized equations of
motion,
\begin{equation}
\ddot{\psi} = \frac{4-h^{2}}{4h^{2}}(\psi_{+} + \psi_{-})
            - \frac{4+h^{2}}{2h^{2}}\psi.
\end{equation}
From this one derives a dispersion relation for small-amplitude travelling
waves, namely,
\begin{equation}
\omega^{2} = \frac{4+h^{2}}{2h^{2}} - \frac{4-h^{2}}{2h^{2}}\cos kh,
\end{equation}
where $k$ is the wave-number. The angular frequency $\omega$
ranges between 1 and $2/h$, the lower bound being responsible for the
threshold velocity.

As the kink travels along the lattice, it hits lattice sites with frequency
$v/h$ sites per unit time. Provided $v \geq h/2\pi $, the kink can excite
radiation of the same frequency, that is $\omega = 2\pi v/h$. However,
if $v<h/2\pi$, then $\omega<1$, and the lattice cannot support such   
radiation. The kink can only excite higher harmonics, so the rate of energy 
dissipation is suddenly cut and the kink velocity becomes ``quasi-stable.''
For $h=1$, this threshold occurs at $v=1/2\pi\approx 0.159$, in good
agreement with the numerical data (fig.\ 7).

A theoretical understanding of the specific shape of the graph is more elusive. 
We make the ad-hoc assumption (motivated by figure 7 of \cite{PK})
that the energy $\Delta E$ lost by a kink in 
traversing a single lattice cell at speed $v$ obeys an exponential 
law:
\begin{equation}
\Delta E = e^{pv-q},
\end{equation}
where $p$ and $q$ are positive constants, properties of the lattice. We further
assume that the kinetic energy of the kink is 
\begin{equation}
E_{K} = \frac{1}{2}cv^{2}
\end{equation}
where $c$ is approximately constant provided $h$ is not too large, as suggested
by the results of section 3 (fig.\ 2). These two equations imply a first order
differential equation for $v(t)$, easily solved to give
\begin{equation}
v(t)=v(0) - A\log(Bt+1),
\end{equation}
where
\begin{eqnarray}
A &=& \frac{1}{p}, \nonumber \\
B &=& \frac{p}{hc}e^{pv(0)-q}. \nonumber
\end{eqnarray}
The dashed curve in fig.\ 7 is a fit of this formula to the numerical data,
taking $A=0.032, B=0.004365, v(0)=0.28$. Estimating $c=1.0986$ by 
averaging
the function  
$f(b)$ for $h=1$, one deduces that $p=31.3$, and $q=17.5$. The fit is good
for velocities greater than the radiation threshold
at $v=0.159$.

\section{Conclusion}

We have described a spatially-discrete sine-Gordon system, which is 
significantly different from the usual (Frenkel-Kontorova) system. There is a
``topological'' lower bound on the kink energy, and an explicit static kink
solution with this energy. The kink moves on a ``level playing-field'': there
is no Peierls-Nabarro potential barrier. In particular, the behaviour of a kink
moving with low speed is much simpler. 

The idea described in section 2 for generating ``topological'' discrete
systems, can be applied to many other continuum theories which have a
Bogomol'nyi bound on the energy of a kink. By way of example, consider the
$\phi^{4}$ equation, where the continuum potential energy density is
$$\frac{1}{4}\left(\frac{\partial\varphi}{\partial x}\right)^{2}
+\frac{1}{4}\left(1-\varphi^{2}\right)^{2}.$$
In equation (4), one replaces $\cos\psi$ by $\frac{1}{3}\varphi^{3}-\varphi$.
The most obvious factorization is then
\begin{eqnarray}
D &=& \Delta\varphi, \nonumber \\
F &=& 1-\frac{1}{3}\left(\varphi_{+}^{2}+\varphi_{+}\varphi+\varphi^{2}
                   \right). \nonumber
\end{eqnarray}
The lattice potential energy is then given by (5) as before, and the kink energy 
is bounded below by $2/3$ (in the kink sector $\varphi\rightarrow\pm 1$ as
$x\rightarrow\pm\infty$). 
               
It would be interesting to investigate breathers and kink-antikink collisions
in this sine-Gordon model.
Again, the absence of the Peierls potential should lead to
behaviour which differs qualitatively from the usual model \cite{BP}. One simple
observation is that in the limiting case $h=2$, there exists a breather
located at a single lattice site. Indeed, the field
\[ \psi(t,x) = \left\{ \begin{array}{ll}
                        0          & \mbox{if $x \neq 0$} \\
                        \theta(t)  & \mbox{if $x=0$}
                        \end{array}
               \right. \]
satisfies the equation of motion if and only if $\theta(t)$ satisfies the
pendulum equation
\[ \ddot{\theta} = -\sin\theta. \]
\\ 
{\bf Acknowledgement:} JMS is supported by a research studentship
awarded by the UK Science and Engineering Research Council.


\end{document}